\documentclass[12pt,a4paper]{article}
\usepackage{amsmath}
\usepackage{amssymb}
\usepackage{appendix}

\newtheorem{theorem}{Theorem}
\newtheorem{acknowledgement}[theorem]{Acknowledgement}

\input{tcilatex}
\begin{document}

\date{}
\title{A model for the reversal of the toroidal rotation in tokamak}
\author{Florin Spineanu and Madalina Vlad \\
National Institute of Laser, Plasma and Radiation Physics\\
Bucharest 077125, Romania}
\maketitle

\begin{abstract}
The transition from toroidal counter- to co- rotation in the core plasma has
been observed at L to H transition in several tokamaks. Spontaneous reversal
has also been observed in TCV beyond a threshold in the density. We develop
a model based on the following phenomenology: (1) the increase of the
gradient of the pressure triggers formation on a fast time scale of cells of
convection (similar to Rayleigh-Benard (RB), but with a single sign of
vorticity); (2) poloidal rotation is induced by the envelope of the
peripheric velocity of the convection cells; via the baroclinic term the
gradients of temperature and density sustain the poloidal rotation against
the decay due to the parallel viscosity; (3) the fast increase of poloidal
flow induces a high time derivative of the radial electric field; (4) the
neoclassical polarization creates a series of parallel accelerations (kiks
on each bounce) of the trapped ions, leading to an increase of the toroidal
precession or to its reversal; the source of energy is the work done by the
radial electric field. (5) the diffusion transfers on resistive scale the
toroidal momentum from the trapped ions to the untrapped ones. The
correlated interactions are examined and the estimated time scales are found
to be compatible with the observations.
\end{abstract}

\section{Introduction}

The spontaneous reversal of the direction of the toroidal rotation in
tokamak has been observed as a change accompanying the L to H transition or
a slow variation of a plasma parameter, like the current or the magnetic
field \cite{Rice0}, \cite{Rice1}, \cite{Rice2}, \cite{Rice3}, \cite{Rice4}.
Spontaneous reversal has also been observed in TCV beyond a threshold in the
density \cite{Bortolon}. A review of these experimental facts can be found
in Refs. \cite{Fiore} and \cite{Rice0a}.

There are several properties of the state of the system that seem to be
correlated with the transition consisting of the flow reversal. There is a
strong radial gradient of temperature which is sustained by external input
of energy into plasma. In general fluid configurations placed under strong
thermal stress [for example in the classical Rayleigh-Benard (RB) system]
the purely conductive state is replaced by a convective regime through
generation of cells of convection \cite{Chandrasekhar}. The transport of
heat and of momentum and angular momentum (via the Reynolds stress) become
much more efficient compared with the diffusive regime. The onset of the
convection structure takes place on short time scale compared with
conduction or flow time scales. The event leading to formation of a regular
convection pattern starts with the generation of isolated hot and cold
streamers (also called \ "plumes" or \ "thermals") that advance into the
fluid environment \cite{Lappa}. The head of the streamer is deformed through
interaction with the static fluid leading to \ "mushroom" shapes also known
from Rayleigh-Taylor experiments. The streamer breaks down leaving vortices
of two signs. These vortices are merging eventually creating the large scale
flow, a mechanism which is universally associated to relaxation \cite%
{KraichnanMontg}, \cite{CorcosSherman}. Under strong temperature gradient
the whole process is very fast.

\bigskip

The presence and/or the possible generation of closed, quasi-coherent
convective flows in the meridional cross section of the tokamak plasma is a
well investigated subject. The particular case of the zonal flows is
included as a subset in this wide class of organized flows (for reviews see 
\cite{DiamondItohs}, \cite{Shukla-PRep}, \cite{Horton1}). Various mechanisms
have been identified to be able to transfer energy \ from the drift wave
turbulence to the flow in a convective structure extending over a
substantial part of the plasma cross section. Modulational instability,
wave-kinetic dynamical transfers in the spectral space, multi-scale
interactions leading to negative viscosity are usually invoked \cite%
{ShuklaPhysRevA}, \cite{DiamondNegVisc}, \cite{ShuklaBalescu}, \cite%
{Shukla-Eur}. A unifying and universal principle is the inverse cascade of
energy in the two-dimensional turbulence \cite{KraichnanMontg}, \cite%
{Shapiro1}, \cite{Pavlenko}, \cite{Weiland1}, \cite{Weiland}. Vortical
structures consisting of robust and coherent flow have been identified in
regimes governed by the Hasegawa-Mima and Hasegawa-Wakatani equations \cite%
{Horton1}. Due to the essential role of the vectorial nonlinearity the space
scales of the these structures is initially at the level of the ion Larmor
radius. The processes or merging lead further to creation of large scale
structure. Another mechanism that has been studied is the tilting
instability of eddies with creation, via Reynolds stress, of flows over
large spatial regions \cite{Rosenbluth}. The structure is usually assumed
with alternate sense of rotation, for several reasons. One of them is the
conservation of the angular momentum integrated over the space occupied by
the cells of convection. This conditions is however imposed to closed
systems, and even for them is not quite stringent \cite{Clercx}. A different
reason comes from assuming the flow as a perturbation, expanding in Fourier
series and retaining the lowest order which by its nature is periodic and
leads to vortices with alternating sense of rotation \cite{Chandrasekhar}.
It also provides compatible flows at the common boundaries of the neighbor
cells. The latter argument should not be seen as a strong restriction if the
distance between the rolls of convection is sufficiently large. Finally we
mention a condition related with the space-time invariances of the
equations, which appears to underlay this choice of the pattern of flow.
Taking as a propotype the Hasegawa-Mima equation one notes the invariance to
reversal of the sign of the vorticity, which implicitely imposes the
presence in the solution of vortices of both signs. However it has been
shown that besides the vectorial nonlinearity the \emph{scalar} or
Korteweg-deVries nonlinearity, of the type $\psi \left( \partial \psi
/\partial y\right) $ where $\psi $ is the streamfunction and $y$ the
poloidal coordinate, is likely to be present in the equation \cite{Nezlin1}, 
\cite{Nezlin2}. Although it does not influence the smaller scales where the
instabilities are developing, it controls the large scales due to the lower
differential degree. This breaks the invariance and allows solutions
consisting of a set of rolls with unique sense of rotation. We argue that
rolls of convection having the same direction of rotating flow are not
excluded by any characteristic of the differential equations governing the
quasi-two-dimensional tokamak plasma and are even more efficient for the
transport of angular momentum in the meridional section. This pattern of
flow has been derived \cite{Shukla} and it arises in many similar
situations. In the science of atmosphere it is known that a distribution of
azimuthal velocity in a vortex can be affected by perturbations leading to
secondary vortices that move around the core of the parent system \cite%
{Schubert}. The geometry of the flow takes the form of a polygonal eye-wall,
which means that a set of smaller vortices are placed symmetrically, with
their centers placed on a circle around the eye-wall axis \cite%
{KossinSchubert}. They all have the same sign of vorticity. Other examples
are accretion disks \cite{accretion} and the Ranque-Hilsch vortex tube \cite%
{Colgate}.

In tokamak there is a poloidal rotation at equilibrium, due to the \emph{%
neoclassical} processes. The poloidal flow is sheared since part of the
plasma cross section is in plateau - banana and part is in Pfirsch-Schluter
regime \cite{HortonNeo}. This just constitutes a \emph{seed} which may favor
a choice of the direction of rotation of the rolls that are going to be
created.

A source of flow that appears to be less investigated is the baroclinic term
arising in the momentum equation of the ion component when it is transformed
by a rotational operator into the equation for toroidally projected
vorticity. It is well suited to exhibit local velocity shear that may
accelerate plasma flow, leading for example to generation of streamers. As
shown below, the streamers constitute a basic agent for building up larger
scale flows.

It is well known that the RB transition from purely conductive to a
convective regime takes place at a threshold, where the buoyancy force
overcomes the viscosity \cite{Chandrasekhar}. In the present case the
threshold has a different nature. The drive force in the early stage is the
baroclinic term, dependent on the main gradients of the density and
temperature, which are supported by the external input. Convective cells are
likely to be created, as shown by numerous investigations (analytical,
numerical, observations). The pattern should obey the symmetries required by
the basic neoclassical rotation which favors formation of rolls all rotating
in the same sense, equivalently, with the same sign of toroidally-directed
vorticity. There is no restriction upon this choice coming from the
conservation of the angular momentum in the meridional plane since momentum
is carried away to the boundary (the system is open). Now, the peripheric
velocities of the chain of large scale vortical cells constitues an
effective poloidal rotation. The magnetic pumping opposes the poloidal
rotation via the parallel viscosity which acts on a time scale determined by
the ion-ion collision frequency \cite{Morris}. This leads us to formulate a
threshold condition: the gradients of density and temperature should be
sufficiently high such as the baroclinic term to sustain the rolls and their
induced poloidal flow against the parallel viscosity damping. The detailed
examination of this threshold condition requires more quantitative analysis
and will not be of the present concern.

\bigskip

We develop a model based on the following phenomenology:

(1) the increase of the gradient of the pressure triggers formation of cells
of convection. The process is similar to the transition from purely
conductive to convective regime in Rayleigh-Benard fluid but the rolls of
convection have a single sign of vorticity; the baroclinic term drives a
fast onset of this pattern of flow;

(2) the peripheric velocities of the convection cells (with unique sense of
rotation) coalesce into an envelope which is equivalent to a poloidal
rotation; in usual circumstances the poloidal rotation is damped by the
parallel viscosity (magnetic pumping) but this is now overcomed by the
Reynolds stress in the regular pattern which is directly sustained by the
gradients of density and temperature. These are, in turn, sustained by the
external input.

(3) the fast increase of poloidal flow induces a high time derivative of the
radial electric field $E_{r}$. The radial momentum balance would allow to
divide the effect of an increase of the $v_{\theta }$ term to the two other
terms, containing the diamagnetic and respectively the toroidal velocities.
However the reaction of the diamagnetic velocity is slower since it depends
on the change of the density profile and the change of the toroidal flow
velocity is mediated by a neoclassical effect.

(4) the neoclassical polarization induced by the variation of the radial
electric field creates a series of parallel accelerations (kick on every
bounce) of the trapped ions, leading to an increase of the toroidal
precession or to its reversal; the source of energy is the work done by the
fast transient growth of $E_{r}$ supported in turn by the pressure gradient
through the convection.

(5) the diffusion transfers on resistive scale the toroidal momentum from
the trapped ions to the untrapped ones, inducing the reversal of direction.

The point (4) roughly describes this situation: on half of the banana (say
in the same direction as the magnetic field $\mathbf{B}$) the trapped ion
feels a certain radial electric field and when it returns on the other half
of the banana orbit (opposite to $\mathbf{B}$) it already is acted upon by a
different radial electric field. The key is that the change of the radial
electric field and the ion bounce on banana take place on comparable time
scales. This is possible due to the fast onset of convection and can be the
explanation of the fast transitions observed in experiments \cite{Burrell1}.

Our model is similar to an \textquotedblleft\ inverse
Ranque-Hilsch\textquotedblright\ effect. In the Ranque-Hilsch vortex (not
fully understood at this time) a strong rotation of a gas leads to
temperature separation \cite{Oliver} and it is supposed that coherent
vortical structures that breaks the azimuthal symmetry of the flow are
responsible for the creation of the gradient of temperature \cite{Colgate}.
In our case we dispose of the strong temperature gradient across the plasma
section and the onset of cells of convection (which break the azimuthal
symmetry) would create the other element, \emph{i.e.} the poloidal flow. The
process is sustained by the heat and density input which is converted into
mechanical stress (radial variation of poloidal rotation) by the cells of
convection. It is transitory with a fast phase (when the reversal occurs)
followed by a slower variation which keeps the time derivative of the radial
electric field at the level required for balancing the parallel viscosity
damping.

We focus in the present work on the logical consistency of the argument,
examining the contributing physical processes under certain simplifications.
The overall result confirms the validity of the model and justifies further
study.

The paper is organised as follows: we explain the role of the baroclinic
term in sustaining the perturbations consisting of hot streamers advancing
along the minor radius. The estimations shows that the time scale of the
convection is very fast. Separately we provide an argument for the poloidal
flow of the density toward the place of origin of the streamer, preventing a
local depression of density and suppression or breaking up of the hot
streamer. This argument is based on another manifestation of the baroclinic
effect, which in this case reveals the possiblity of the instability called
Chaplygin gas, or anomalous polytropic. The next section is devoted to
demonstrate that the peripheric velocity of the rolls of convection leads to
an effective poloidal flow. The argument is here geometric and on this basis
one can easily superpose physical processes like Reynolds stress and
collisional diffusion of the momentum in the vecinity of the rolls of
convection. The last part consists of simply adapting the arguments of
Hinton and Robertson \cite{HintonRobertson} to show that the neoclassical
polarization induces an acceleration of the toroidal motion of the bananas.

\section{The hot streamers in the early phase of the onset of the convection
rolls}

We consider a perturbation in the two-dimensional section of a tokamak
plasma, consisting of the radial displacement of an element of plasma from
the radius $r_{0}$ to higher radii, \emph{i.e.} to toward the edge. For fast
motion the element preserves its temperature and advances as a stream of hot
plasma to higher $r$. The Figure 1 shows the geometry of the perturbation
fixing also the system of reference $\left( x\equiv r,y\equiv r\theta
,z\right) $. The smooth surface represents the equilibrium pressure profile
on which it is superposed the perturbation related to the advancing hot
streamer. The curved arrow shows the effect of the baroclinic term. The two
arrows at the right of the perturbations have the following meaning: the one
directed against the perturbation represents the $\mathbf{\nabla }T$ of the
hot stream relative to the environment. The one poining approximately in the
inverse $x\equiv r$ direction is the gradient of the equilibrium density (or
mass), $\mathbf{\nabla }\rho $.

The density at the radius where the streamer begins its motion is $n\left(
r_{0}\right) $ and the temperature at the same location is $T\left(
r_{0}\right) $. The displacement of the hot plasma element along the radius
is $\delta r=r-r_{0}$ which is positive in our system. On this distance the
equilibrium (background) pressure has varied and has the value%
\begin{equation}
p\left( r\right) =p\left( r_{0}\right) +\frac{dp}{dr}\delta r=p\left(
r_{0}\right) -p\left( r\right) \frac{1}{\left\vert L_{p}\right\vert }\delta r
\label{eq100}
\end{equation}%
where $L_{p}<0$ on an usual equilibrium pressure profile. This $p\left(
r\right) $ is the pressure in the environment plasma at a distance $\delta r$
from the point of initialization of the thermal streamer. Assuming that
there is no loss of heat via diffusion the hot stream will have at the same
position the pressure $p\left( r_{0}\right) $. The gradient due to this
contrast between the hot streamer and the environment is directed (along
poloidal direction at $r$) from the boundary of the thermal streamer to its
axis. We will choose to look at the left side of the hot streamer which
means that the gradient of the pressure is directed along the positive $y$
axis, $\widehat{\mathbf{e}}_{\theta }$. The width of the hot stream is of
the order of the radius of the roll, $R$ and the gradient of pressure
transversal on the axis of the streamer is%
\begin{equation}
\frac{\delta p}{R}=\eta \frac{p\left( r_{0}\right) -p\left( r\right) }{R}%
=\eta p\left( r\right) \frac{1}{\left\vert L_{p}\right\vert }\delta r\frac{1%
}{R}  \label{eq101}
\end{equation}%
In order to keep a realistic set of assumptions we must consider the loss of
heat from the streamer via conduction to the environment plasma. An
elementary option is to introduce an arbitrary nondimensional small factor $%
\eta $ and consider from now on the fraction $\eta $ of the perturbation of
the pressure. Now, the gradient of the perturbed pressure interacts with the
equilibrium gradient of the density $\rho _{0}$ to give the baroclinic term%
\begin{eqnarray}
\frac{1}{\rho _{0}^{2}}\mathbf{\nabla }\rho _{0}\times \mathbf{\nabla }p &=&%
\frac{1}{\rho _{0}^{2}}\left\vert \mathbf{\nabla }\rho _{0}\right\vert
\left( -\widehat{\mathbf{e}}_{r}\right) \times \left\vert \mathbf{\nabla }%
p\right\vert \left( \widehat{\mathbf{e}}_{\theta }\right)  \label{eq102} \\
&=&\frac{1}{\rho _{0}^{2}}\left\vert \mathbf{\nabla }\rho _{0}\right\vert
\left\vert \mathbf{\nabla }p\right\vert \left( -\widehat{\mathbf{e}}%
_{z}\right)  \notag
\end{eqnarray}%
and using Eq.(\ref{eq101}) we have 
\begin{equation}
\frac{1}{\rho _{0}^{2}}\mathbf{\nabla }\rho _{0}\times \mathbf{\nabla }p=%
\frac{1}{m_{i}n\left( r\right) }\frac{1}{\left\vert L_{n}\right\vert }\eta 
\frac{p\left( r\right) }{\left\vert L_{p}\right\vert }\frac{\delta r}{R}%
\left( -\widehat{\mathbf{e}}_{z}\right)  \label{eq103}
\end{equation}%
The two quantities: the depth of penetration of the streamer $\delta r$ and
the radius of the cell of convection $R$ produced by the roll-up of the
streamer have similar magnitudes and we approximate%
\begin{equation}
\frac{\delta r}{R}\sim 1  \label{eq104}
\end{equation}%
The baroclinic term is directed along the negative $z$ (toroidal) axis and
enhances the local gradients of the velocity of the streamer: the velocity
on the axis of the hot streamer is accelerated. With this specification of
the direction of the enhancement of the velocity by the time-derivative of
the local vorticity, we will only take from now on its absolute value 
\begin{equation}
\left\vert \frac{\partial \omega }{\partial t}\right\vert \sim \frac{1}{%
\delta t}\frac{v}{R}\sim \frac{1}{m_{i}n\left( r\right) }\frac{1}{\left\vert
L_{n}\right\vert }\eta \frac{p\left( r\right) }{\left\vert L_{p}\right\vert }
\label{eq105}
\end{equation}%
Here $v\equiv v_{r}$ is the speed of the hot streamer, along its axis. The
interval of time $\delta t$ will later be taken a characteristic turn-over
of the convection roll. The gradient length of the equilibrium pressure is%
\begin{equation}
\frac{1}{L_{p}}=\frac{1}{p}\frac{\partial p}{\partial r}=\frac{1}{nT}\left( T%
\frac{\partial n}{\partial r}+n\frac{\partial T}{\partial r}\right) =\frac{1%
}{L_{n}}+\frac{n}{p}\frac{\partial T}{\partial r}  \label{eq106}
\end{equation}%
with both terms having the same sign, negative along the increasing minor
radius direction. From Eqs.(\ref{eq105}) and (\ref{eq106}) we have the
estimation 
\begin{equation}
\frac{1}{\delta t}\frac{v}{R}\sim \eta \frac{T}{m_{i}}\left( \frac{1}{%
\left\vert L_{n}\right\vert ^{2}}+\frac{1}{\left\vert L_{n}\right\vert }%
\frac{1}{\left\vert L_{T}\right\vert }\right)  \label{eq107}
\end{equation}%
Now, if we assume that the gradient length of the density is much larger
than $\left\vert L_{T}\right\vert $, we will retain only the second term in
the paranthesis%
\begin{equation}
\frac{1}{\delta t}\frac{v}{R}\sim \eta \frac{1}{m_{i}}\frac{1}{\left\vert
L_{n}\right\vert }\left\vert \mathbf{\nabla }T\right\vert  \label{eq108}
\end{equation}%
Taking%
\begin{equation}
\delta t\sim \frac{R}{v}  \label{eq109}
\end{equation}%
we obtain%
\begin{equation}
v^{2}\sim R^{2}\eta \frac{1}{m_{i}\left\vert L_{n}\right\vert }\left\vert 
\mathbf{\nabla }T\right\vert  \label{eq110}
\end{equation}%
The gradient of the equilibrium temperature at the position and on the
spatial extension of the convection roll must be a fraction of the gross
estimation $T/R$ and we introduce an arbitrary nondimansional small
parameter $\lambda $ 
\begin{equation}
\left\vert \mathbf{\nabla }T\right\vert \sim \lambda \frac{T}{R}
\label{eq111}
\end{equation}%
it results%
\begin{equation}
v^{2}\sim R^{2}\eta \lambda \frac{1}{m_{i}\left\vert L_{n}\right\vert }\frac{%
T}{R}=\eta \lambda \frac{R}{\left\vert L_{n}\right\vert }v_{i,th}^{2}
\label{eq112}
\end{equation}%
Except for the two small parameters $\eta $ and $\lambda $ this formula
shows that the speed of rotation at the periphery of the roll is very high,
comparable with the ion thermal velocity. Including a rough estimation for
the conductive loss (similar to the Rayleigh-Benard case) by the factor $%
\eta =0.1$ and the variation of the equilibrium temperature over the cell's
extension, by the factor $\lambda =0.1$, we still get a high value $v\sim
0.1\times v_{th,i}$.

\bigskip

\section{The local perturbation of the density along the poloidal direction}

\subsection{The instability of the Chaplygin-type or negative pressure \ }

The perturbation is initiated at a radius $r_{0}$, which means that the
density in a small region around the circle $r_{0}$ will flow toward the
point (say $\theta _{0}$) where the streamer originates. There the poloidal
aflux is converted into radial flow of the hot streamer. This radial flow is
accelerated by the baroclinic effect and from the Eq.(\ref{eq112}) we see
that the speed can reach high values. A local depression of density can
occur if the rate of filling with density the foot of the streamer takes
place at a rate which is slower than that with which it is removed by the
streamer. This is an important aspect which we examine in the following. The
continuity equation with only the velocity along the streamer is%
\begin{equation}
\frac{\partial n}{\partial r}+\frac{\partial }{\partial r}\left(
v_{r}n\right) =0  \label{eq113}
\end{equation}%
or%
\begin{equation}
\frac{\partial }{\partial t}\ln n+\frac{\partial v_{r}}{\partial r}+v_{r}%
\frac{\partial }{\partial r}\ln n=0  \label{eq114}
\end{equation}%
The radial velocity $v_{r}$ is zero for radii smaller than $r_{0}$ and rises
to an almost constant value on a small radial distance $\frac{\partial v_{r}%
}{\partial r}\sim v_{r}\delta \left( r-r_{0}\right) $ . The equation of
continuity becomes%
\begin{equation}
\frac{d}{dt}\ln n=-v_{r}\delta \left( r-r_{0}\right)  \label{eq115}
\end{equation}%
where $\frac{d}{dt}=\frac{\partial }{\partial t}+v_{r}\frac{\partial }{%
\partial r}$. The solution is%
\begin{equation}
\ln n\left[ r\left( t\right) \right] =\ln n\left[ r\left( t_{0}\right) %
\right] -\int_{t_{0}}^{t}dt^{\prime }v_{r}\delta \left[ r\left( t\right)
-r_{0}\right]  \label{eq116}
\end{equation}%
where the integral along the trajectory is the inverse of the Lagrangian
derivation $d/dt$. The expression shows that the density is simply carried
along the streamer, with a discontinuity in the small region at the foot of
the streamer. This leads to a depression of density if the neighbor regions
are not providing aflux of particles. We will show that there is an
instability consisting of strong modulation of the density on the poloidal
circle $r_{0}$ which can lead to accumulation of density in periodic
poloidal positions. Although this instability and the formation of the
streamer are independent, they can act concurrently, avoiding suppression of
the streamer or its breaking up with further propagation as blob. This is
the reason for which we examine now the poloidal modulation of the density
when a conjunctural factor (the streamer) imposes a flow along $\theta $.

Assuming incompressibility $\mathbf{\nabla \cdot v}=0$ in the small region
around the streamer's foot we have $\delta v_{\theta }/\delta l_{\theta
}=\delta v_{r}/\delta l_{r}$ with comparable variation scales $\delta
l_{\theta }\sim \delta l_{r}$. The two velocities, $v_{\theta }$ and $%
v\equiv v_{r}$ are therefore related $v_{\theta }\sim v$. Accordingly, the
acceleration of the streamer velocity $v$ along the radius induces an
accelaration of the flow $v_{\theta }$ on the poloidal circle.%
\begin{equation}
\frac{\partial v}{\partial t}\sim \frac{\partial v_{\theta }}{\partial t}
\label{eq117}
\end{equation}%
which simply means that the time variation of the velocity in the streamer
is also the time variation of the poloidal dynamics.

From now on our considerations will be refering to the poloidal direction
only, more precisely to a narrow region around the circle of radius $r_{0}$
where the hot streamer is born. As said before the acceleration on the
radial direction of the flow in the hot streamer induces an acceleration of
the aflux of particles, along the poloidal circle, $\partial v_{\theta
}/\partial t$, toward the foot of the streamer. In this region the dynamics
along the hot streamer, essentially radial, is coupled to the dynamics along
the poloidal circle. If the main characteristic of the hot streamer is the
contrast of temperature relative to the environment in which it advances,
the dynamics along the poloidal region regards essentially the density. This
is because the temperature on the circle is constant while the particles
must flow toward the streamer 's foot to sustain the streamer flux and this
generates density perturbation. We introduce the density $\rho $ on the
circle and we have that in the region of perturbation induced by the
streamer, $\rho =\rho _{0}$. The problem becomes one-dimensional, along the
circle $r_{0}$, with the coordinate $y\equiv r_{0}\theta $ and we take the
origin of $y$ at the axis of the streamer, \emph{i.e.} $y=0$ at $\theta _{0}$%
. For uniformity, we change the notation, $v_{y}\equiv v_{\theta }$.

A new manifestation of the baroclinic effect takes place along the poloidal
direction. We first remind the case of the streamer, where the baroclinic
force arises from the contrast of temperature between the hot streamer and
the environment, in combination with the gradient of the background density.
Now the baroclinic effect arises from the perturbation of density along the
poloidal direction, in combination with the gradient of the equilibrium
pressure, $\mathbf{\nabla }p_{0}$, in which the gradient of the equilibrium
temperature $T_{0}\left( r\right) $ is active. Assume that due to the flux
of particles towards the foot of the streamer $y=0$ there is a perturbation
in the form of a lump of density at this point,. We have, focusing our
considerations on the left side of the hot streamer, $\partial \rho
/\partial y>0$, 
\begin{equation}
\frac{\partial \mathbf{\omega }}{\partial t}\sim \frac{1}{\rho ^{2}}\mathbf{%
\nabla }\rho \times \mathbf{\nabla }p_{0}=\frac{1}{\rho ^{2}}\left( \frac{%
\partial \rho }{\partial y}\right) \left\vert \frac{dp_{0}}{dr}\right\vert 
\widehat{\mathbf{e}}_{z}  \label{eq118}
\end{equation}%
The effect of this baroclinic force is to enhance the flow along $y$ toward $%
y=0$. Taking a distance on $y$ of perturbation of $\rho \left( y\right) $ of
the scale as the radius of the rolls of convection, $R$, we have an
estimation of the effect on velocity coming from the left to $y=0$, the axis
of the streamer. There is acceleration%
\begin{equation}
\frac{\partial v_{y}}{\partial t}\sim R\frac{\partial \omega }{\partial t}=R%
\left[ \frac{1}{\rho ^{2}}\frac{\partial \rho }{\partial y}\left\vert \frac{%
\partial p_{0}}{\partial r}\right\vert \right] =R\frac{1}{\rho }\frac{%
\partial \rho }{\partial y}\left\vert \frac{\partial \left(
T_{0}/m_{i}\right) }{\partial r}\right\vert  \label{eq119}
\end{equation}%
and the balance of forces along $y$ is%
\begin{equation}
\rho \left( \frac{\partial v_{y}}{\partial t}+v_{y}\frac{\partial v_{y}}{%
\partial y}\right) =\rho \left( R\frac{1}{\rho }\frac{\partial \rho }{%
\partial y}\left\vert \frac{\partial \left( T_{0}/m_{i}\right) }{\partial r}%
\right\vert \right)  \label{eq120}
\end{equation}%
We set%
\begin{equation}
\sigma \equiv R\left\vert \frac{\partial \left( T_{0}/m_{i}\right) }{%
\partial r}\right\vert =R\left\vert \frac{\partial v_{th,i}^{2}}{\partial r}%
\right\vert  \label{eq121}
\end{equation}%
with dimensions of squared velocity.

We will express the term in the right hand side as a derivation to $y$, such
as to exhibit the structure of the force resulting from the gradient of a
scalar \emph{pressure}, which we will have to identify. This means to find a
function $F\left( y\right) $ such as 
\begin{equation}
\rho \left( \sigma \frac{1}{\rho }\frac{\partial \rho }{\partial y}\right)
\equiv -\frac{\partial }{\partial y}F\left( \rho \right) =-\frac{dF\left(
\rho \right) }{d\rho }\frac{\partial \rho }{\partial y}  \label{eq122}
\end{equation}%
We obtain%
\begin{eqnarray}
\frac{dF\left( \rho \right) }{d\rho } &=&-\sigma  \label{eq123} \\
F\left( \rho \right) &=&-\sigma \rho  \notag
\end{eqnarray}%
and this allows us to characterize the term arising from the \emph{%
baroclinic }effect as a force produced by the $y$-gradient of a pressure%
\begin{equation}
\rho \left( \sigma \frac{1}{\rho ^{2}}\frac{\partial \rho }{\partial y}%
\right) \equiv -\frac{\partial }{\partial y}\overline{p}  \label{eq124}
\end{equation}%
where the equivalent pressure, as a scalar function whose gradient is the
force, is identified as%
\begin{equation}
\overline{p}=-\sigma \rho  \label{eq125}
\end{equation}%
and is \emph{negative}. The equations of motion for an element of plasma
situated on the poloidal circle at radius $r_{0}$ consists of the
deformation displacements in the poloidal direction due to the force
generated from the gradient of the scalar pressure%
\begin{equation}
\frac{\partial v_{y}}{\partial t}+v_{y}\frac{\partial v_{y}}{\partial y}=-%
\frac{1}{\rho }\frac{\partial }{\partial y}\overline{p}  \label{eq126}
\end{equation}%
and the continuity equation%
\begin{equation}
\frac{\partial \rho }{\partial t}+\frac{\partial }{\partial y}\left( \rho
v_{y}\right) =0  \label{eq127}
\end{equation}

The same problem has been treated by Ott \cite{Ott} where the temperature is 
\emph{negative}. The pressure in the present case is \emph{negative} with 
\emph{linear} dependence on the perturbation of the density. This is the
typical $m=1$ case of the general treatment of Trubnikov and Zhdanov \cite%
{Trubnikov}.

In the Appendix A we show, following Trubnikov and Zhdanov \cite{Trubnikov},
how this system is solved. The solution has the following characteristic:
the density on the poloidal circle has regions of broad, quasi-uniform
magnitude separated by very high (spiky) accumulations. In the actual
physical process the strong accumulations coincide with the initialization
of the hot streamers and effectively support the flow of density they
represent.

\subsection{Comment on the theory of the anomalous polytropic gas}

We comment on the applicability of the results obtained from the model of
anomalous polytropic gas to the physical problem of breaking of the
azimuthal symmetry of the density and flow in tokamak with generation of
radial streamers and further of rolls of convection.

We have two distinct processes: (1) a hot stream of plasma leaving the
region at radius $r_{0}$ and advancing in the colder environment at higher $%
r $, sustained by a baroclinic force. The baroclinic force is generated by
the combined effect of the contrast of temperature between the streamer and
the environment and the gradient of equilibrium density profile. It is a
fast process and a density depletion around $r_{0}$ can appear. (2) an
instability of the density profile along the poloidal direction at $r_{0}$,
again sustained by a baroclinic effect. This is determined by the
combination between the gradient of the perturbed density along $\theta $
and the gradient of the equilibrium pressure. The effect is generation of
high accumulation of density at periodic positions along the poloidal
direction.

These two processes are interacting: the tendency of accumulation of density
at certain location on the poloidal circle supports the flow in the
streamers. The decrease of the density at $r_{0}$ by the streamer flow is
compensated by the aflux of density associated to the (Chaplygin gas - type)
instability of the density on $\theta $. Observations and numerical
simulations suggest that the streamers ( "plumes" or \ "thermals") of hot
fluid inhibit the instability on the transversal direction by the high rate
of radial flow which sometimes lead to breaking-up of streamer from its
original region and separation as a blob.

\bigskip

The system of equations describing the instability in the early phase can be
reduced to an $1D$ model of a gas with anomalous polytropic (the pressure is
negative and $\overline{p}\sim -\sigma \rho $). The solution consists of
regions of broad and smooth variation of the variables $\left( \rho
,v\right) $ and very localised, spiky regions of high concentration of
density.

The instability of Chaplygin gas is fast: the streamers (which appear in the
model as concentrations of density) are formed on a time scale given by the
equation%
\begin{equation}
\frac{\Lambda }{\tau }\sim \sigma ^{1/2}  \label{eq12701}
\end{equation}%
which results from the eigenmodes determined in the linearized system (the
dispersion relation already shows this characteristic scales). Here $\Lambda 
$ and $\tau $ are typical wavelength and time scales of the Chaplygin
instability on $y$. Let us take $\rho _{0}=m_{i}n\sim 10^{-8}\ kg/m^{3}$, $%
R\sim a/10\sim 0.1\ \left( m\right) $, 
\begin{equation}
\frac{dp_{0}}{dr}\sim \frac{1}{\left( a/10\right) }nT=3\times 10^{4}\ \left( 
\frac{J}{m^{4}}\right)  \label{eq12702}
\end{equation}%
for $n=10^{19}\ \left( part/m^{3}\right) $ and $T=2000\ \left( eV\right) $.
Then%
\begin{equation}
\sigma ^{1/2}\approx 5\times 10^{5}\ \left( \frac{m}{s}\right)
\label{eq12703}
\end{equation}%
Now we estimate the distance between the centres of two neighbor rolls,
taking the pattern consisting of $m=5$ convective rolls with their centres
at $r_{0}\sim a/3$, for a minor radius $a=1\ \left( m\right) $, $\Lambda
=2\pi r_{0}/m=0.4\ \left( m\right) $. Then the time scale is%
\begin{equation}
\tau =\frac{\Lambda }{\sigma ^{1/2}}\sim 10^{-6}\ \left( s\right)
\label{eq12704}
\end{equation}%
which is very fast. Moreover, this time scale is compatible with the time
scale determined independently from the flow of the hot streamer, since from
Eq.(\ref{eq109}) we also obtain $\tau \sim 10^{-6}...10^{-5}\ \left(
s\right) $.

\bigskip

\section{The poloidal velocity as the envelope of the rotations in the cells
of convection (rolls)}

The peripheric velocity of the convection rolls is transferred to plasma at
larger $r$ both by Reynolds stress and by collisional drag. On the other
hand even the geometry of the flow around closed rolls leads to a continuous
(but wavy) streamlines which can represent a continuous poloidal flow. We
want to show how to obtain the envelope representing the connected poloidal
flow arising from separated rolls having rotations in the same direction.
Therefore we will restrict the treatment to the pure geometric properties of
the ideal flow. The Reynolds stress and collisions will enhance the
efficiency of producing a directed poloidal flow.

The treatment is based on complex formalism \cite{Crowdy1} for the plane
flow. We consider the \emph{fluid} in the complex $z$ plane with a set of
vortices placed at different positions. We need the \emph{complex potential}
of the flow in the $z$ plane, $w\left( z\right) $. The problem is not solved
directly in the physical plane of the flow but in a plane $\zeta $ connected
with $z$ by a \emph{conformal transformation}. The idea is to take in the $%
\zeta $-plane a structure which is standard (\emph{canonical}). It consists
of (1) the circle with the center in $\zeta =0$ and unit radius $\left\vert
\zeta \right\vert =1$ and (2) a set of $M$ circles with centres $\delta _{j}$
and radii $q_{j}$ placed inside the circle $\left\vert \zeta \right\vert =1$.

By the \emph{conformal map} $\zeta \rightarrow z\left( \zeta \right) $ this
corresponds to $M+1$ fixed circular obstacles in the $z$ plane of the
physical flow. Our rolls of convection are patches of rotating fluid for
which it is only specified the \emph{circulation} of the fluid around (\emph{%
i.e.} the integral of the tangential velocity on the closed contour) $\gamma
_{j}\ \ ,\ \ j=1,M$. The fluid moves in the $z$ plane with the condition of
keeping the given value $\gamma _{j}$ of the circulation at every circle $j$%
. This does not uniquely specify the flow. One needs the condition at
infinity, far from the regions where the circular vortices are given.

After defining the geometry it is possible to construct the \emph{complex
potential} in the $\zeta $ plane. It is easily shown that it is

\begin{equation}
w\left( \zeta \right) =-\frac{i}{2\pi }\ln \left[ \frac{\zeta -\alpha }{%
\left\vert \alpha ^{\ast }\right\vert \left( \zeta -\frac{1}{\alpha ^{\ast }}%
\right) }\right] -\frac{i}{2\pi }\ln \zeta +\text{const}  \label{eq128}
\end{equation}%
This function represents the flow around a unit vortex located in $\alpha
\in \zeta -$plane and is constant on the circle bounding the large domain $%
\left\vert \zeta \right\vert =1$. The \emph{complex potential} gives a
circulation of $-1$ around the circle $\left\vert \zeta \right\vert =1$ , $%
\doint\limits_{\left\vert \zeta \right\vert =1}\mathbf{v\cdot dl}=-1$ where $%
\mathbf{v}$ is calculated from $w$ according to the formulas of the Appendix
B.

To simplify, we take the geometry in the $z$-plane of the physical flow as
consisting of three convection rolls represented by three disks with centers
at $z_{-1}=-d$, $z_{0}=0$, $z_{+1}=+d$, with equal radii, $s$. The
circulation around each vortex roll is $\gamma $.

\bigskip

In the plane $\zeta $ one has a circle of radius $\left\vert \zeta
\right\vert =1$. This will be transformed by the conformal map $\zeta
\rightarrow z$ in the circle with center at $z_{0}=0$, which is one of the
fixed physical vortices (or rolls of convection), the central one. Two other
circles are inside this $\left\vert \zeta \right\vert =1$ domain, located at 
$\delta _{-1}=-\delta $, $\delta _{+1}=+\delta $ and having equal radii $=q$%
. Now the geometry of the $\zeta $ plane is defined. One has to write the 
\emph{complex potential} in the $\zeta $ plane.

To impose a particular circulation on the contour of the $j^{th}$ circular
patch of vorticity (or obstacle) from the set of $M$ in the $z$-plane, one
defines in the $\zeta $-plane the potential%
\begin{equation}
G_{j}\left( \zeta ,\alpha \right) =-\frac{i}{2\pi }\ln \left( \frac{\omega
\left( \zeta ,\alpha \right) }{\left\vert \alpha \right\vert \omega \left(
\zeta ,\theta _{j}\left( \frac{1}{\alpha ^{\ast }}\right) \right) }\right) \
\ \text{for}\ \ j=1,...,M  \label{eq129}
\end{equation}%
The function that has been introduced is%
\begin{equation}
\theta _{j}\left( \zeta \right) =\delta _{j}+\frac{q_{j}^{2}\zeta }{1-\delta
_{j}^{\ast }\zeta }  \label{eq130}
\end{equation}%
is a Mobius map. Then $G_{j}\left( \zeta ,\alpha \right) $ is the \emph{%
complex potential} of a flow in the plane $\zeta $ generated by a vortex of
unit strength $+1$ in the point $\zeta =\alpha $ and making a circulation $%
-1 $ around the circle $\delta _{j}$ of radius $q$. This potential has \emph{%
zero circulation} around any of the other $M-1$ circles in the $\zeta $
plane and around the circle $\left\vert \zeta \right\vert =1$. By
superposition $-\dsum\limits_{j=0,j=1,...,M}\gamma _{j}G_{j}\left( \zeta
,\beta \right) $ , one obtains in the plane $\zeta $ the \emph{complex
potential} for the canonical geometry: $\left\vert \zeta \right\vert =1,\
\delta _{-1},\delta _{+1}$. Since we assume that the convection rolls have
equal circulations $\gamma _{j}=\gamma \ \ $for$\ \ j=1,2,3$, we obtain $%
w\left( \zeta \right) =-\gamma _{-}G_{-}\left( \zeta ,\alpha \right) -\gamma
_{0}G_{0}\left( \zeta ,\alpha \right) -\gamma _{+}G_{+}\left( \zeta ,\alpha
\right) $. This \emph{complex potential }is in $\zeta $ plane and from here
it must be carried back onto the physical ($z$) plane, by a \emph{conformal
transformation} which requires numerical calculation. The resulting flow
have streamlines that show formation of poloidal rotation as we depart
progressively from the line where the centers of the vortices are located 
\cite{Crowdy1}. Since this is a gemetrical property it means that the
poloidal flow occurs simultaneously with the onset of convection rolls, 
\emph{i.e.} on a very fast time scale.

\section{The effect of the fast increase of the radial electric field and
the acceleration of the toroidal flow of bananas}

The high density and temperature gradients (or both), supported by external
sources, lead to formation of large scale cells of convection. Due to the
background neoclassical rotation the rolls have all the same direction of
rotation. The baroclinic term induces high peripheric velocities in the
rolls and the envelope of these peripheric flows is a poloidal rotation. The
rise of the poloidal velocity takes place on the same time scale as the
onset of the convection rolls and this is governed by the baroclinic term
acting on the initial hot stream. It is a fast process, similar to the onset
of convection in the classical Rayleigh-Benard system. One important aspect
should be underlined: the flow in the convection rolls (hence the poloidal
rotation) is sustained by the gradients of the equilibrium temperature
and/or the equilibrium density, which in turn are sustained by external
sources. Therefore the flow in the rolls of convection (implicitely the
induced poloidal velocity) is sustained by external sources. This completely
changes the usual picture on the poloidal rotation, the damping due to the
magnetic pumping is now balanced and overcommed by the external sources.
This also suggests the existence of a threshold, which will be examined in a
separate work.

The dynamics in the radial direction consists of charge separation and has
weak effect on the force balance along this direction. We consider the
approximate radial equilibrium of forces%
\begin{equation}
E_{r}\approx \frac{1}{\left\vert e\right\vert n}\frac{dp}{dr}+v_{T}B_{\theta
}-v_{\theta }B_{T}  \label{eq131}
\end{equation}%
as valid over the interval of time increase of $v_{\theta }$. In Appendix C
we argue that the fast change of $v_{\theta }$ is mainly reflected in a
change of the radial electric field $E_{r}$, on the same time scale, leading
to the approximative balance%
\begin{equation}
\frac{\partial E_{r}}{\partial t}\approx -B_{T}\frac{\partial v_{\theta }}{%
\partial t}  \label{eq132}
\end{equation}
Since we have%
\begin{equation}
\frac{v_{th,i}}{Rq}\lesssim \frac{\partial }{\partial t}\ll \Omega _{ci}
\label{eq133}
\end{equation}%
it results that the time change of the radial electric field will modify the
motion of the bananas.

We are in the regime described by Hinton and Robertson \cite{HintonRobertson}
where during the motion of the particle on the banana orbit there is
effective change of the radial electric field which acts on the particle. In
the Ref.\cite{HintonRobertson} the effect is found to be manifested in two
ways. First, there is a steady radial motion of the bananas. Second, in each
bounce period, there is an acceleration of the motion in the toroidal
direction. The effects are due to the fact that (schematically) the particle
feels a certain radial electric field on half of its orbit and a different
radial electric field when it returns along the second half of the orbit.

The main assumption that has led us to this estimation is that%
\begin{equation}
\frac{\partial \omega }{\partial t}=\frac{v}{R}\frac{1}{\delta t}
\label{eq134}
\end{equation}%
\emph{i.e.} the baroclinic term acts all around the circular path of the
rolled up flow in the cell and accelerates the rotation. The time scale is
given by the time of rotation in the cell.%
\begin{equation}
\delta t\sim \frac{R}{v}\approx \frac{\left( R\left\vert L_{n}\right\vert
\right) ^{1/2}}{\left( \eta \lambda \right) ^{1/2}v_{i,th}}  \label{eq135}
\end{equation}

The time variation of the poloidal velocity is the time of acceleration%
\begin{equation}
\frac{v}{\delta t}\sim \eta \lambda \frac{v_{i,th}^{2}}{\left\vert
L_{n}\right\vert }  \label{eq136}
\end{equation}%
and the time derivative of the radial electric field is%
\begin{equation}
\frac{\partial E_{r}}{\partial t}\approx \frac{\partial }{\partial t}\left(
-v_{\theta }B_{T}\right) \sim -\eta \lambda \frac{v_{i,th}^{2}}{\left\vert
L_{n}\right\vert }B_{T}  \label{eq137}
\end{equation}

At every bounce the banana gets a toroidal acceleration \cite%
{HintonRobertson}%
\begin{equation}
\delta u=\left( \frac{\partial E_{r}}{\partial t}\right) \frac{1}{B_{\theta }%
}\delta t_{bounce}  \label{eq138}
\end{equation}%
where%
\begin{equation}
\delta t_{bounce}=\frac{qR_{0}}{v_{i,th}}  \label{eq139}
\end{equation}%
\begin{equation}
\delta u\sim -\eta \lambda \frac{v_{i,th}^{2}}{\left\vert L_{n}\right\vert }%
B_{T}\frac{1}{B_{\theta }}\frac{qR_{0}}{v_{i,th}}=-v_{i,th}\eta \lambda 
\frac{B_{T}}{B_{\theta }}\frac{qR_{0}}{\left\vert L_{n}\right\vert }
\label{eq140}
\end{equation}%
This kick of acceleration, leading to the change of the toroidal velocity of
the trapped particle $\delta u$ is due to the variation of the radial
electric field during the time interval the trapped ion makes a complete
bounce. The number of bounces that a trapped particle makes in an interval
of time $\Delta t$ is%
\begin{equation}
n=\frac{\Delta t}{\delta t_{bounce}}=\frac{\Delta t}{qR_{0}/v_{i,th}}
\label{eq141}
\end{equation}%
During the time interval $\Delta t$ the total change of the toroidal
velocity of a trapped particle is%
\begin{equation}
\Delta u=n\delta u=-\Delta tv_{i,th}^{2}\frac{1}{\left\vert L_{n}\right\vert 
}\eta \lambda \frac{B_{T}}{B_{\theta }}  \label{eq142}
\end{equation}%
This change is substantial. On a time interval of one millisecond $\Delta
t=10^{-3}s$, with $L_{n}\approx 5\ m$ , $B_{T}/B_{\theta }\sim 5$, $%
v_{i,th}=10^{5}\ m/s$ and taking as before $\eta =0.1$, $\lambda =0.1$ the
change of the toroidal velocity of a single trapped particle is of the order
of the thermal velocity. When the initial toroidal flow is opposite to this
acceleration there is reversal of rotation. The energy requested for this
change is provided, in our model, by the equilibrium gradients of
temperature and density.

\section{Discussion and summary}

There are other mechanisms that can sustain or modify the spontaneous
toroidal rotation. The drift wave turbulence creates a statistical ensemble
of fluctuating velocities and the coupling of the radial and poloidal
components can have a non-zero correlation (Reynolds stress) with spatial
dependence. The divergence of this correlation is a driving force for the
rotation. Due to the long correlation length in the toroidal direction, the
Reynolds stress is efficient in the poloidal plane and in particular is able
to induce zonal flows. If it overcomes the magnetic damping it can induce
poloidal rotation and so generate a non-vanishing $\partial E_{r}/\partial t$
and the subsequent neoclassical polarization. The latter modifies the
toroidal precession, \emph{i.e.} the plasma toroidal rotation. A detailed
analysis of this connection should be made, with focus on transitory
processes, as observed in Alcator C-Mode \cite{Rice0}. However few aspects
can be discussed qualitatively. Since it is based on a statistical ensemble\
the Reynolds stress needs a strong poloidal-symmetry breaking mechanism
otherwise the average is zero. Indeed the symmetry breaking can exist,
analogous to the flows arising in the Rayleigh-Benard experiments after
higher order bifurcations (flows are called \ "winds"). However their
manifestation is slower that all the other time scales: conduction,
convection. It is difficult to see how the onset of a poloidal flow via only
Reynolds stress can take place on time scales less than milliseconds, if the
background consists of random drift waves. On the other hand it is well
established that large scale convective flows have much higher Reynolds
stress than the $2D$ quasi-isotropic turbulence. This is partly the reason
for the present investigation, where we precede any effect of change of the
rotation by the onset of a chain of convection cells. We further note one
aspect that results from the combined effect of convective cells and the
Stringer drive and compare it with the Reynolds stress. A chain of
convective cells (with either unidirectional or alternating vorticity
direction) induces a periodic modulation of the local rate of transport. The
neoclassical Stringer effect converts this modulation into an effective
poloidal torque, possibly higher than the damping due to magnetic pumping.
In a similar case but in the absence of convection cells \cite%
{McCarthyHassam}\textbf{, }the ratio of the magnitude of the Reynolds stress
over the Stringer drive was found small. The presence of convective cells
enhances both the Reynolds stress and the Stringer drive. Further study is
necessary for a correct comparison of magnitudes.

\bigskip

We have proposed a phenomenological model for the reversal of the toroidal
rotation in tokamak. The model assembles effects that separately are known
to appear in tokamak plasma and have been studied both experimentally and
theoretically. The generation of convection structures with flows that are
quasi-coherent on large scales in the meridional section is in general
admitted as a particularly effective way of transport of angular momentum
and of energy across the magnetic field. We consider the situation of cells
of convection with the same sense of rotation flow, disposed around the
magnetic axis in the region extending from the core plasma to the confining
zone, a structure that is compatible with the basic neoclassical flow. It is
also inspired by atmospheric and protoplanetary flows where such
distribution occurs in hydrodynamic regimes very similar to the one in
tokamak. The convection is sustained by equilibrium gradients of density and
temperature (which in turn are fed by external sources) and this drives the
poloidal velocity which is generated by the combined, unidirectional,
peripheric flows of the cells. Then the decay due to parallel viscosity
(magnetic pumping) can be overcomed. The time scales are short due to the
baroclinic term which accounts for the generation of vorticity. The fast
increase of the poloidal velocity induces a fast variation of the radial
electric field, on a time scale that is comparable with the bounce time. On
part of the banana trajectory the ions see a magnitude of the radial
electric field and on rest of the trajectory they see a different magnitude.
This produces both radial motion and the change of velocity in toroidal
direction. The collisions will spread this drive to the rest of the ion
population.

The principal merits of this theoretical explanation is that it provides a
connection between poloidal and toroidal rotation. Usually the direction of
influence is reversed and based of instabilities: the sheared parallel flow
induces instabilities whose turbulent Reynolds stress acts to generate and
sustain the poloidal rotation against the magnetic pumping decay \cite%
{DongHorton}, \cite{SuYushHorton}. This mechanism is active in situation
where the turbulence is known to induce poloidal flows acting as internal
transport barriers, while the basic flow is observed to be toroidal. It can
interfere with the generation of convective structures, for example in
formation of Goertler vortices \cite{Drazin}, etc. In the present case the
poloidal flow is a geometric consequence of the flow in the convection
structures. The driving effect of the poloidal flow on the toroidal flow has
a fast transient phase which is decissive since it involves neoclassical
polarization on the time scale of the bounce on bananas.

\bigskip

The different components of this phenomenology appear to be compatible and
to combine in a structure which is logically coherent. Future work is needed
for a more detailed examination of this theory.

\begin{acknowledgement}
Work supported partially by the Contracts BS-2 and BS-14 of the Association
EURATOM - MEdC Romania. The views presented here do not necessarly represent
those of the European Commission.
\end{acknowledgement}

\bigskip

\begin{appendices}
\section{Appendix. The Chaplygin gas instability}
\renewcommand{\theequation}{A.\arabic{equation}} \setcounter{equation}{0}

\subsection{The formalism for the Chaplygin gas instability}

We will linearize the equations (\ref{eq126}) and (\ref{eq127}), \emph{i.e.}%
\begin{equation}
\frac{\partial \rho }{\partial t}+\frac{\partial }{\partial y}\left( \rho
v_{y}\right) =0  \label{a2_100}
\end{equation}%
and%
\begin{equation}
\frac{\partial v_{y}}{\partial t}+v_{y}\frac{\partial v_{y}}{\partial y}=-%
\frac{1}{\rho }\frac{\partial }{\partial y}\overline{p}  \label{a2_101}
\end{equation}%
around the state $\rho =\rho _{0}$ ,$\ v_{y}=0$ and obtain%
\begin{eqnarray}
\frac{\partial }{\partial t}\left( \rho _{0}+\widetilde{\rho }\right) +\frac{%
\partial }{\partial y}\left[ \left( \rho _{0}+\widetilde{\rho }\right) 
\widetilde{v}_{y}\right]  &=&0  \label{a2_102} \\
\frac{\partial \widetilde{v}_{y}}{\partial t}+\widetilde{v}_{y}\frac{%
\partial \widetilde{v}_{y}}{\partial y} &=&-\frac{1}{\rho _{0}+\widetilde{%
\rho }}\frac{\partial }{\partial y}\left[ -\sigma \widetilde{\rho }\right]  
\nonumber
\end{eqnarray}%
We have%
\begin{eqnarray}
\frac{\partial \widetilde{\rho }}{\partial t}+\rho _{0}\frac{\partial 
\widetilde{v}_{y}}{\partial y} &=&0  \label{a2_103} \\
\frac{\partial \widetilde{v}_{y}}{\partial t} &=&-\frac{1}{\rho _{0}}\left(
1-\frac{\widetilde{\rho }}{\rho _{0}}\right) \frac{\partial }{\partial y}%
\left( -\sigma \widetilde{\rho }\right) \approx \frac{1}{\rho _{0}}\sigma 
\frac{\partial \widetilde{\rho }}{\partial y}  \nonumber
\end{eqnarray}%
Now we take $\widetilde{\rho }\sim \exp \left[ i\left( ky-\omega t\right) %
\right] $ and $\widetilde{v}_{y}\sim \exp \left[ i\left( ky-\omega t\right) %
\right] $ and inserting in the linearized equations we get%
\begin{equation}
-i\omega \widetilde{\rho }+\rho _{0}ik\left[ -\frac{1}{i\omega }\frac{\sigma 
}{\rho _{0}}ik\widetilde{\rho }\right] =0  \label{a2_104}
\end{equation}%
or%
\begin{equation}
\omega =\pm ik\left( \sigma \right) ^{1/2}  \label{a2_105}
\end{equation}%
The positive imaginary part indicates instability, similar to the
Rayleigh-Taylor case.

The\ numerical solution of the system of equations \cite{Ott} %
 shows that there are broad regions of weak variation and very peaked
accumulations, which evolve to singular spatial spikes.

The equations in Ref.\cite{Trubnikov} page 143 are relevant to the present
case, for our \emph{negative} pressure%
\begin{equation}
\overline{p}=-\sigma \rho   \label{a2_106}
\end{equation}%
If we introduce this expression in the equation of motion we obtain%
\begin{equation}
\frac{\partial v_{y}}{\partial t}+v_{y}\frac{\partial v_{y}}{\partial x}=-%
\frac{1}{\rho }\frac{\partial }{\partial y}\overline{p}=\frac{1}{\rho _{0}+%
\widetilde{\rho }}\frac{\partial }{\partial y}\left( \sigma \widetilde{\rho }%
\right)   \label{a2_107}
\end{equation}%
and the system can be rewritten in terms of two variables $\widetilde{\rho }%
\ \ $and$\ \ v_{y}$. 
\begin{eqnarray}
\frac{\partial \widetilde{\rho }}{\partial t}+\rho _{0}\frac{\partial v_{y}}{%
\partial y}+\frac{\partial }{\partial y}\left( \widetilde{\rho }v_{y}\right)
&=&0  \label{a2_108} \\
\frac{\partial v_{y}}{\partial t}+v_{y}\frac{\partial v_{y}}{\partial y} &=&%
\frac{\sigma }{\rho _{0}}\frac{\partial }{\partial y}\widetilde{\rho } 
\nonumber
\end{eqnarray}

The equations belong to the \emph{elliptic} type and therefore they do not
have running waves. These equations correspond to the case $m=1$ in the
general treatment made by Trubnikov and Zhdanov and it is shown that they lead
to the Laplace equation.

We now try to apply the \emph{hodograph transformation} to the equations of
our case. Instead of the variables $\widetilde{\rho }\left( y,t\right) $ and 
$v_{y}\left( y,t\right) $ \ \ we will use the \emph{inverted} variables $%
y\left( \widetilde{\rho },v_{y}\right) $ and $t\left( \widetilde{\rho }%
,v_{y}\right) $ with the relationships between the derivatives%
\begin{eqnarray}
\frac{\partial v_{y}}{\partial t} &=&\frac{1}{w}\frac{\partial y}{\partial 
\widetilde{\rho }}\ \ ,\ \ \frac{\partial v_{y}}{\partial y}=-\frac{1}{w}%
\frac{\partial t}{\partial \widetilde{\rho }}  \label{a2_109} \\
\frac{\partial \widetilde{\rho }}{\partial t} &=&-\frac{1}{w}\frac{\partial y%
}{\partial v_{y}}\ \ ,\ \ \frac{\partial \widetilde{\rho }}{\partial y}=%
\frac{1}{w}\frac{\partial t}{\partial v_{y}}  \nonumber
\end{eqnarray}%
The Jacobian of the transformation is%
\begin{equation}
w=\frac{\partial y}{\partial \widetilde{\rho }}\frac{\partial t}{\partial
v_{y}}-\frac{\partial y}{\partial v_{y}}\frac{\partial t}{\partial 
\widetilde{\rho }}  \label{a2_110}
\end{equation}

Now we substitute these equations in the system, to make the transformation
to the inverted variables $\left( y,t\right) $ instead of $\left( \widetilde{%
\rho },v_{y}\right) $.%
\begin{eqnarray}
&&\frac{\partial \widetilde{\rho }}{\partial t}+\rho _{0}\frac{\partial v_{y}%
}{\partial y}+\frac{\partial }{\partial y}\left( \widetilde{\rho }%
v_{y}\right)   \label{a2_111} \\
&=&-\frac{1}{w}\frac{\partial y}{\partial v_{y}}+\rho _{0}\left( -\frac{1}{w}%
\frac{\partial t}{\partial \widetilde{\rho }}\right) +\frac{1}{w}\frac{%
\partial t}{\partial v_{y}}v_{y}+\widetilde{\rho }\left( -\frac{1}{w}\frac{%
\partial t}{\partial \widetilde{\rho }}\right) =0  \nonumber
\end{eqnarray}%
We separate the derivative $\partial y/\partial v_{y}$%
\begin{equation}
\frac{\partial y}{\partial v_{y}}=-\left( \rho _{0}+\widetilde{\rho }\right) 
\frac{\partial t}{\partial \widetilde{\rho }}+v_{y}\frac{\partial t}{%
\partial v_{y}}  \label{a2_112}
\end{equation}

Making the same substitution in the second equation we have%
\begin{eqnarray}
\frac{\partial v_{y}}{\partial t}+v_{y}\frac{\partial v_{y}}{\partial y} &=&%
\frac{\sigma }{\rho _{0}}\frac{\partial }{\partial y}\widetilde{\rho }
\label{a2_113} \\
\frac{1}{w}\frac{\partial y}{\partial \widetilde{\rho }}+v_{y}\left( -\frac{1%
}{w}\frac{\partial t}{\partial \widetilde{\rho }}\right)  &=&\frac{\sigma }{%
\rho _{0}}\frac{1}{w}\frac{\partial t}{\partial v_{y}}  \nonumber
\end{eqnarray}%
and we separate the derivative $\partial y/\partial \widetilde{\rho }$ from
this equation%
\begin{equation}
\frac{\partial y}{\partial \widetilde{\rho }}=v_{y}\frac{\partial t}{%
\partial \widetilde{\rho }}+\frac{\sigma }{\rho _{0}}\frac{\partial t}{%
\partial v_{y}}  \label{a2_114}
\end{equation}

Now we have two partial derivatives of $y$ to the two new variables $\left( 
\widetilde{\rho },v_{y}\right) $. We impose the condition of compatibility
of the mixed second order derivatives%
\begin{equation}
\frac{\partial }{\partial \widetilde{\rho }}\frac{\partial y}{\partial v_{y}}%
=\frac{\partial }{\partial v_{y}}\frac{\partial y}{\partial \widetilde{\rho }%
}  \label{a2_115}
\end{equation}%
which leads to%
\begin{equation}
\frac{1}{\rho _{0}+\widetilde{\rho }}\frac{\partial }{\partial \widetilde{%
\rho }}\left[ \left( \rho _{0}+\widetilde{\rho }\right) ^{2}\frac{\partial t%
}{\partial \widetilde{\rho }}\right] +\frac{\sigma }{\rho _{0}}\frac{%
\partial ^{2}t}{\partial v_{y}^{2}}=0  \label{a2_116}
\end{equation}

To go further (and show that this equation leads to the Laplace equation) we
make another substitution of variables%
\begin{equation}
r\equiv \left( \rho _{0}+\widetilde{\rho }\right) ^{1/2}\ \ ,\ \ z\equiv 
\frac{1}{2}\frac{v_{y}}{\sqrt{\sigma /\rho _{0}}}  \label{a2_117}
\end{equation}%
Then%
\begin{equation}
\rho _{0}+\widetilde{\rho }=r^{2}\ \ ,\ \ v_{y}=\alpha z  \label{a2_118}
\end{equation}%
where $\alpha \equiv 2\sqrt{\sigma /\rho _{0}}$ and the change of variables
implies%
\begin{eqnarray}
\frac{\partial }{\partial r} &=&\frac{\partial }{\partial \widetilde{\rho }}%
\frac{\partial \widetilde{\rho }}{\partial r}=2r\frac{\partial }{\partial 
\widetilde{\rho }}  \label{a2_119} \\
\frac{\partial }{\partial z} &=&\frac{\partial }{\partial v_{y}}\frac{%
\partial v_{y}}{\partial z}=\alpha \frac{\partial }{\partial v}  \nonumber
\end{eqnarray}%
or%
\begin{equation}
\frac{\partial }{\partial \widetilde{\rho }}=\frac{1}{2r}\frac{\partial }{%
\partial r}\ \ ,\ \ \frac{\partial }{\partial v_{y}}=\frac{1}{\alpha }\frac{%
\partial }{\partial z}  \label{a2_120}
\end{equation}%
Inserting in the equation, we have%
\begin{equation}
\frac{1}{\rho _{0}+\widetilde{\rho }}\frac{\partial }{\partial \widetilde{%
\rho }}\left[ \left( \rho _{0}+\widetilde{\rho }\right) ^{2}\frac{\partial t%
}{\partial \widetilde{\rho }}\right] +\frac{\sigma }{\rho _{0}}\frac{%
\partial ^{2}t}{\partial v_{y}^{2}}=0  \label{a2_121}
\end{equation}%
which can be rewritten%
\begin{equation}
\frac{\partial ^{2}t}{\partial r^{2}}+\frac{3}{r}\frac{\partial t}{\partial r%
}+\frac{\partial ^{2}t}{\partial z^{2}}=0  \label{a2_122}
\end{equation}

Following the general procedure (of Trubnikov and Zhdanov) we make the substitution%
\begin{equation}
\psi \equiv rt  \label{a2_123}
\end{equation}%
and replace the variable $t$ by $\psi $%
\begin{eqnarray}
t &=&\frac{\psi }{r}  \label{a2_124} \\
\frac{\partial t}{\partial r} &=&-\frac{1}{r^{2}}\psi +\frac{1}{r}\frac{%
\partial \psi }{\partial r}  \nonumber \\
\frac{\partial ^{2}t}{\partial r^{2}} &=&\frac{2}{r^{3}}\psi -\frac{1}{r^{2}}%
\frac{\partial \psi }{\partial r}-\frac{1}{r^{2}}\frac{\partial \psi }{%
\partial r}+\frac{1}{r}\frac{\partial ^{2}\psi }{\partial r^{2}}  \nonumber
\end{eqnarray}%
and we obtain%
\begin{equation}
\frac{\partial ^{2}\psi }{\partial r^{2}}+\frac{1}{r}\frac{\partial \psi }{%
\partial r}-\frac{1}{r^{2}}\psi +\frac{\partial ^{2}\psi }{\partial z^{2}}=0
\label{a2_125}
\end{equation}

This is the Laplace equation for a potential of the form%
\begin{equation}
\Psi \left( r,\varphi ,z\right) =\psi \left( r,z\right) \cos \varphi 
\label{a2_126}
\end{equation}

\subsection{The solution of the Laplace equation}

In Ref.\cite{Trubnikov} it is explained the so-called \ "Evolutionary
principle ", a restriction placed on the solutions of the Laplace equation:
the physically useful solutions must satisfy the condition that, returning
in time to the infinite past $t\rightarrow -\infty $ the function $\Psi $ or 
$\left( \rho ,v\right) $ must take the values of the equilibrium profile.
This means $\rho \rightarrow \rho _{0}$ and $v_{y}\rightarrow 0$, or,
equivalently, $r=1$ and $z=0$. In accord with the last form of the Laplacean
expressed in variables $\left( r,\varphi ,z\right) $ this corresponds to a
circle or radius $1$ in the plane $z=0$. All the initial conditions at
infinite past must be taken on a thin region around this circle. Since the
Laplace equation also describes the electrostatic field, Trubnikov and Zhdanov
exploit this analogy by formulating the problem as the solution of an
electrostatic problem in space, generated by a certain distribution of
charges on the circle $r=1$ and $z=0$. The solutions that are admissible
must be singular on this circle. The distribution of charges is equivalently
described in terms of multipoles.  Trubnikov and Zhdanov show that only two
multipoles are useful for applications: the \emph{Coulomb} distribution and
the \emph{dipole} distribution of charges on the circle. The first provides
periodic solutions in real (\emph{i.e.} physical) space. The\ "$2$%
-multipole" (dipole) generates three types of solutions in the physical
space: the \emph{hill}, the \emph{well} and the doublet consiting of \emph{%
well+hill}.

The solution of the Laplace equation in the space $\left( r,\varphi
,z\right) $ with boundary conditions given on the circle $\left(
r=1,z=0\right) $ requires to assume the toroidal geometry and therefore to
introduce the system of coordines%
\begin{equation}
r=\lambda \sinh \xi \ \ ,\ \ z=\lambda \sin \eta   \label{a2_127}
\end{equation}%
where $\lambda =\left( \cosh \xi +\cos \eta \right) ^{-1}$ and the metric is%
\begin{equation}
dr^{2}+dz^{2}=\left( \lambda d\xi \right) ^{2}+\left( \lambda d\eta \right)
^{2}  \label{a2_128}
\end{equation}%
The Lam\'{e} coefficients are $h_{\xi }=\lambda $ , $h_{\varphi }=r=\lambda
\sinh \xi $ and $h_{\eta }=\lambda $. The Laplace equation is%
\begin{equation}
\frac{\partial }{\partial \xi }\left( \lambda \sinh \xi \frac{\partial \Psi 
}{\partial \xi }\right) +\frac{\partial }{\partial \eta }\left( \lambda
\sinh \xi \frac{\partial \Psi }{\partial \eta }\right) +\frac{\lambda }{%
\sinh \xi }\frac{\partial ^{2}\Psi }{\partial \varphi ^{2}}=0  \label{a2_129}
\end{equation}

In our particular case, which corresponds to $m=1$ of the general theory, we
must seek solutions of the type $\Psi =\psi \left( r,z\right) \cos \varphi $
and after the substitution 
\begin{equation}
\psi \left( r,z\right) =r^{-1/2}f  \label{a2_130}
\end{equation}%
the amplitude is written as a series of terms with \emph{separated variables}%
\begin{equation}
f=\dsum\limits_{n=0}^{\infty }f_{n}\left( \xi \right) \cos \left( n\eta
\right)   \label{a2_131}
\end{equation}

Now it is introduced a new variable instead of $\xi $%
\begin{equation}
\chi \equiv \coth \xi   \label{a2_132}
\end{equation}%
and this transforms the Laplace equation into the equation for Legendre
polynomials%
\begin{equation}
\frac{d}{d\chi }\left[ \left( 1-\chi ^{2}\right) \frac{df_{n}}{d\chi }\right]
+\left[ \nu \left( \nu +1\right) -\frac{n^{2}}{1-\chi ^{2}}\right] f_{n}=0
\label{a2_133}
\end{equation}%
for $\nu \equiv -3/2$. The solutions are, in general%
\begin{equation}
P_{\nu }^{n}\left( \xi \right) \ \ \text{and}\ \ Q_{\nu }^{n}\left( \xi
\right)   \label{a2_134}
\end{equation}%
However only the solutions $Q_{\nu }^{n}\left( \xi \right) $ must be
retained since they have the property to become singular on the circle $r=1$
at $t\rightarrow -\infty $.

The solution is characterised by large regions where the function is smooth
with weak variation and local spikes which evolve in time to singularity. The
distribution of density is then compatible with the independent creation of
streams of plasma which require strong density flow.
\end{appendices}

\begin{appendices}
\section{Appendix. The complex potential of flow of a system of circular
vorticity patches}
\renewcommand{\theequation}{B.\arabic{equation}} \setcounter{equation}{0}

The flow is described by a complex function named \emph{complex potential} $%
w $. It is composed from:

\begin{enumerate}
\item the real part, $\varphi \left( x,y\right) $, is the potential of the
flow, \emph{i.e.} the function from which the components of the velocity are
derived%
\begin{equation}
v_{x}=\frac{\partial \varphi }{\partial x}\ \ ,\ \ v_{y}=\frac{\partial
\varphi }{\partial y}  \label{a1_100}
\end{equation}%
This is the simple expression of the physical fact that the flow is \emph{%
potential}, \emph{i.e. }it is derived as the gradient of the scalar function 
$\mathbf{v=\nabla }\varphi $. The flow is therefore \emph{irrotational}, the
vorticity is zero $\mathbf{\omega }=\mathbf{\nabla \times }\left( \mathbf{%
\nabla }\varphi \right) \equiv \mathbf{0}$.

\item the imaginary part, $\psi \left( x,y\right) $ which is the \emph{%
streamfunction} and is related to the potential function $\varphi $ by the
fact that it also gives the components of the velocity%
\begin{equation}
v_{x}=-\frac{\partial \psi }{\partial y}\ \ ,\ \ v_{y}=\frac{\partial \psi }{%
\partial x}  \label{a1_101}
\end{equation}%
Then%
\begin{equation}
\frac{\partial \varphi }{\partial x}=-\frac{\partial \psi }{\partial y}\ \
,\ \ \frac{\partial \varphi }{\partial y}=\frac{\partial \psi }{\partial x}
\label{a1_103}
\end{equation}%
which are the Riemann-Cauchy conditions for holomorphy of the \emph{complex
potential} $w\left( x,y\right) $. The lines of \emph{streamfunction} $\psi
\left( x,y\right) $ are perpendicular on the lines of the potential $\varphi 
$. The velocity is tangent to the lines $\psi \left( x,y\right) =$const.
\end{enumerate}

Then the complex potential is%
\begin{eqnarray}
w\left( x,y\right)  &=&\varphi \left( x,y\right) +i\psi \left( x,y\right) \
\ \text{or}  \label{a1_104} \\
w\left( z\right)  &=&\varphi \left( z\right) +i\psi \left( z\right)  
\nonumber
\end{eqnarray}%
Since $\varphi $ and $\psi $ are the real and imaginary part of a
holomorphic function (in a region of holomorphicity) both are \emph{harmonic}
functions%
\begin{eqnarray}
\Delta \varphi  &=&0  \label{a1_105} \\
\Delta \psi  &=&0\ \ \text{in the regions outside vortices}  \nonumber
\end{eqnarray}%
and the flow has no vorticity with the exception of the closed regions where
the circulations is defined, $\gamma _{j}$.
\end{appendices}

\begin{appendices}
\section{Appendix. The fast increase of the poloidal velocity}
\renewcommand{\theequation}{C.\arabic{equation}} \setcounter{equation}{0}

The problem consists of calculation of the radial polarization current when
the poloidal velocity is strongly forced. The process has two stages: the
first is the onset of the poloidal rotation, as envelope of the peripheric
flows of a chain of convection cells in the meridional plane; the second is
the sustainement of the poloidal rotation against magnetic damping and
viscosity. The first stage is very fast, as is suggested by the similarity
with the Rayleigh-Benard bifurcation conduction/convection and as is
confirmed by the evaluation of the drive of the streamers by the baroclinic
term in the vorticity equation. In this stage the rise of the poloidal flow
generates a time-varying radial polarization electric field which acts on
trapped ions on the time scale of their bounce producing a toroidal
acceleration. The work done by the radial electric field on trapped ions is
supported by the energy coming from the flow in the convection cells and
this one is in turn supported by gradients of equilibrium plasma parameters
(ultimetely from external input). 

In general the radial projection of the
ion momentum connects terms that contain the poloidal velocity $v_{y}^{i}$,
the toroidal velocity $v_{z}^{i}$, the radial electric field $E_{r}$ and the
diamagnetic term $\sim dp_{i0}/dr$. Since now the poloidal velocity is
forced, it is important to see which of the other terms will get the maximum
effect on the short time scale of the onset of the convection rolls. Based
on reasonable assumptions we show that the drive of $\partial
v_{y}^{i}/\partial t$ goes primarly to the radial electric field ($\partial
E_{r}/\partial t$, the polarization). This further justifies to look to its
effect on trapped ions. 

The estimations of the important effects will proceed along (and it is part
of) the derivation of the initial value problem for the poloidal velocity, as shown in Ref. \cite{FlMa}
The radial current results from the Ampere's law%
\begin{equation}
\left. \mathbf{\nabla \times B}\right\vert _{x}=0=\mu _{0}j_{x}+\varepsilon
_{0}\mu _{0}\frac{\partial E_{x}}{\partial t}  \label{118}
\end{equation}%
where 
\begin{equation}
j_{x}\left( t\right) =-env_{x}^{e}+env_{x}^{i}  \label{jx1}
\end{equation}

We use a {\it slab} geometry where the coordinates are: $x\equiv r\;\;$with$\;\;\widehat{\mathbf{e}}_{x}\equiv -%
\widehat{\mathbf{e}}_{r}\;\;$(toward the plasma interior), $y\equiv r\theta
\;\;$with$\;\;\widehat{\mathbf{e}}_{y}\equiv \widehat{\mathbf{e}}_{\theta
}\;\;$(poloidal), and $z\equiv R\varphi \;\;$with$\;\;\widehat{\mathbf{e}}%
_{z}\equiv \widehat{\mathbf{e}}_{\varphi }\;\ $(toroidal). The conservation
of the ion momentum in the radial direction is 
\begin{equation}
n_{i}m_{i}\left( \frac{\partial v_{x}^{i}}{\partial t}+v_{y}^{i}\frac{\partial
v_{x}^{i}}{\partial y}\right) =-\frac{\partial p_{i0}}{\partial x}%
+en_{i}E_{x}+en_{i}v_{y}^{i}B_{z}-en_{i}v_{z}^{i}B_{y}+R_{x}  \label{119}
\end{equation}%
The lines of flow in the poloidal direction are wiggly due to the discrete
chain of rolls but on the average the term with the poloidal $\left(
y\right) $ derivative of the radial $v_{x}^{i}$ velocity vanishes. In the
following we we neglect the resistive terms. 
\begin{equation}
v_{y}^{i}=\frac{m_{i}}{eB_{z}}\frac{\partial v_{x}^{i}}{\partial t}-\frac{%
E_{x}}{B_{z}}+\frac{B_{y}}{B_{z}}v_{z}^{i}+\frac{1}{n_{i}m_{i}\Omega _{i}}%
\frac{\partial p_{i0}}{\partial x}  \label{120}
\end{equation}%
the last term being the diamagnetic flow contribution. The equation for the
poloidal velocity $v_{y}^{i}$ is indicative for the effect that can result
from the fast time variation, when the poloidal flow is strongly forced by
convection rolls. The time variation of $v_{y}^{i}$ can induce a time
variation of the diamagnetic term, but this can be considered small since
the change of the equilibrium profile of the density involves diffusive time scale.
 The change of the density due to the perturbation (\emph{%
i.e.} the convection cells) appears here in second order and will be
neglected.

Further, we have to bear in mind our main objective, which is to show the
possibility of coupling of the poloidal and toroidal rotations via the
neoclassical polarization, acting on trapped ions. For this reason we can
postpone the inclusion of the direct ({\emph i.e.} not mediated by neoclassical polarization) 
effect of $\partial v_{y}^{i}/\partial t
$ upon the term $\frac{B_{y}}{B_{z}}v_{z}^{i}$. The relative importance of this term
will be discussed at the end of this Appendix. Then we will use the
approximations%
\begin{equation}
v_{y}^{i}\approx -\frac{E_{x}}{B_{z}}+\frac{m_{i}}{eB_{z}}\frac{\partial
v_{x}^{i}}{\partial t}  \label{151}
\end{equation}

\bigskip 

The conservation of the ion momentum in the poloidal direction is 
\begin{equation}
n_{i}m_{i}\left( \frac{\partial v_{y}^{i}}{\partial t}+v_{x}^{i}\frac{%
\partial v_{y}^{i}}{\partial x}\right) =-\frac{\partial p_{i0}}{\partial y}%
+en_{i}E_{y}-en_{i}v_{x}^{i}B_{z}+R_{y}  \label{122}
\end{equation}%
Since this calculation is made on the points along the poloidal circle where
the convection rolls are closed the radial ion velocity $v_{x}^{i}$ on this
circle and the nonlinear term $v_{x}^{i}\frac{\partial v_{y}^{i}}{\partial x}
$ are small. Also the poloidal $\left( y\right) $ components of the electric
field $E_{y}$ and of the gradient of the pressure are vanishingly small
compared with the rest of the terms (actually they are neoclassical
corrections related to trapping and toroidal viscosity, which are here
comparatively small). It results%
\begin{equation}
v_{x}^{i}=-\frac{1}{\Omega _{i}}\left( \frac{\partial v_{y}^{i}}{\partial t}%
\right) \left[ 1+\frac{1}{\Omega _{i}}\left( \frac{\partial v_{y}^{i}}{%
\partial x}\right) \right] ^{-1}  \label{124}
\end{equation}

Now we use the Ampere's law%
\begin{equation}
-en_{e}v_{x}^{e}=-en_{i}v_{x}^{i}-\varepsilon _{0}\frac{\partial E_{x}}{%
\partial t}  \label{1516}
\end{equation}%
In this expression we insert the radial electric field from Eq.(\ref{151})
and the radial ion's velocity from Eq.(\ref{124}). We obtain a single
equation for the poloidal ion velocity 
\begin{eqnarray}
-env_{x}^{e} &=&\frac{en}{\Omega _{i}}\left( \frac{\partial v_{y}^{i}}{%
\partial t}\right) \frac{1}{U}+\varepsilon _{0}B_{z}\left( \frac{\partial
v_{y}^{i}}{\partial t}\right)   \label{154} \\
&&+\frac{\varepsilon _{0}m_{i}}{e\Omega _{i}}\left( \frac{\partial
^{3}v_{y}^{i}}{\partial t^{3}}\right) \frac{1}{U}-\frac{2\varepsilon
_{0}m_{i}}{e\Omega _{i}^{2}}\left( \frac{\partial ^{2}v_{y}^{i}}{\partial
t^{2}}\right) \left( \frac{\partial ^{2}v_{y}^{i}}{\partial x\partial t}%
\right) \frac{1}{U^{2}}  \nonumber \\
&&-\frac{2\varepsilon _{0}m_{i}}{e\Omega _{i}^{3}}\left( \frac{\partial
v_{y}^{i}}{\partial t}\right) \left( \frac{\partial ^{2}v_{y}^{i}}{\partial
x\partial t}\right) ^{2}\frac{1}{U^{3}}-\frac{\varepsilon _{0}m_{i}}{e\Omega
_{i}^{2}}\left( \frac{\partial v_{y}^{i}}{\partial t}\right) \left( \frac{%
\partial ^{3}v_{y}^{i}}{\partial x\partial t^{2}}\right) \frac{1}{U^{2}} 
\nonumber
\end{eqnarray}%
with the notation $U\equiv 1+\frac{1}{\Omega _{i}}\left( \frac{\partial
v_{y}^{i}}{\partial x}\right) $. For the phase of the acceleration the
variation on $x$ (radial) is not relevant and a simplified form is obtained

\begin{equation}
-en_{e}v_{x}^{e}=\varepsilon _{0}B_{z}\left( 1+\frac{c^{2}}{v_{A}^{2}}%
\right) \frac{\partial v_{y}^{i}}{\partial t}+\varepsilon _{0}\frac{m_{i}^{2}%
}{e^{2}B_{z}}\frac{\partial ^{3}v_{y}^{i}}{\partial t^{3}}  \label{153}
\end{equation}%
The calculations must be repeated for electrons, to derive the expression of 
$v_{x}^{e}$. The relevant terms have however factors of $\Omega _{e}^{-1}$
and since $\Omega _{e}\gg \Omega _{i}$ the electron current can be neglected 
\begin{equation}
\left( \varepsilon _{\perp }/\varepsilon _{0}\right) \frac{\partial v_{y}^{i}%
}{\partial t}+\frac{1}{\Omega _{i}^{2}}\frac{\partial ^{3}v_{y}^{i}}{%
\partial t^{3}}\approx 0  \label{1537}
\end{equation}%
The quantity $\varepsilon _{\perp }=\varepsilon _{0}\left(
1+c^{2}/v_{A}^{2}\right) $ is the effective plasma dielectric in the
direction transversal to the strong magnetic field. The initial condition is the
acceleration of the poloidal flow resulting from forcing at the onset of the
convection, $K\equiv dV_{y}/dt$. As said before, this process (similar to RB
bifurcation) is expected to be very fast but its details are still beyond
analytical description.%
\[
\left( \frac{1}{\Omega _{i}^{2}}\frac{\partial ^{2}}{\partial t^{2}}+\frac{%
\varepsilon _{\perp }}{\varepsilon _{0}}\right) \frac{\partial v_{y}^{i}}{%
\partial t}+K\Theta \left( t\right) =0
\]%
where $\Theta \left( t\right) $ is the Heaviside function. With the time
scaled $t\rightarrow t\Omega _{c}$ the equation becomes 
\[
\overset{\cdot \cdot }{w}+bw=-\overline{K}\Theta \left( t\right) 
\]%
Here $w\left( t\right) \equiv \partial v_{y}^{i}/\partial t$ and the dot
means derivative to the scaled time variable. The notations are $%
b=\varepsilon _{\perp }/\varepsilon _{0}$ and $\overline{K}\equiv \Omega
_{c}^{-1}K$. The solution, obtained by Laplace transform \cite{FlMa} is%
\[
w\left( t\right) =\frac{-\overline{K}}{\varepsilon _{\perp }/\varepsilon _{0}%
}+\left( w_{0}-\frac{-\overline{K}}{\varepsilon _{\perp }/\varepsilon _{0}}%
\right) \cos \left[ \left( \varepsilon _{\perp }/\varepsilon _{0}\right)
^{1/2}t\right] +w_{1}\left( \frac{1}{\varepsilon _{\perp }/\varepsilon _{0}}%
\right) ^{1/2}\sin \left[ \left( \varepsilon _{\perp }/\varepsilon
_{0}\right) ^{1/2}t\right] 
\]%
For simplicity we have not included the terms that induce the decay of the oscillations \cite{FlMa}.
The initial conditions are given for the first and the second derivatives of
the velocity, respectively $w_{0}=w\left( t=0\right) $, and $w_{1}=\overset{%
\cdot }{w}\left( t=0\right) $. From the definition of $K$ we have $%
w_{0}=w\left( t=0\right) =\Omega _{c}^{-1}K=\frac{1}{\Omega _{i}}\left( 
\frac{dV_{y}}{dt}\right) _{t=0}$ . The non-oscillatory part of  acceleration 
$\overline{w}\left( t\right) =\frac{-\overline{K}}{\varepsilon _{\perp
}/\varepsilon _{0}}$ results from estimations made in the Section 2.
During the onset of convection cells the poloidal velocity reach values
comparable to the ion thermal velocity $\sim 10^{5}\ \left( m/s\right) $ in
the characteristic time of advancement of the streamer, $10^{-6}\ \left(
s\right) $. Then the initialization is  $\left( dV_{y}/dt\right) _{t=0}\sim
10^{11}\ \left( m/s^{2}\right) $ and the non-oscillatory part has the order
of magnitude   
\begin{equation}
\left( \frac{dV_{y}}{dt}\right) _{t=0}\frac{1}{\varepsilon _{\perp
}/\varepsilon _{0}}\sim 10^{8}\ \left( m/s^{2}\right)   \label{140}
\end{equation}%
Now we can return to Eq.(\ref{120}) and use this estimation to compare the
terms that are supposed to share the effect of the fast increase of the
poloidal velocity. The only term that may preoccupate contains the toroidal
velocity. Taking as reference the experimental observation \cite{Rice1} the
toroidal velocity is of the order of $10^{4}\ \left( m/s\right) $ and any
change takes place on a time scale of fractions of a second. This means that
the observed toroidal acceleration is substantially slower than Eq.(\ref{140}%
). Besides, this term is multiplied by the small ratio $B_{y}/B_{z}\ll 1$.
We conclude that the fast time variation of the poloidal velocity can only
be followed by the radial electric field, which justifies Eq.(\ref{eq132}).
\end{appendices}

\bigskip

\bigskip

\bigskip

\textbf{Figure 1.}

\textbf{Figure caption}: The system of reference and the geometry of the
perturbation. The smooth surface represents the equilibrium pressure profile
on which the temperature perturbation of the hot streamer is superposed. The
curved arrow indicates the direction of the baroclinic velocity
acceleration. The two arrows at the right side of the perturbation
represents the gradients: the arrow directed to the left is $\mathbf{\nabla }%
T$ of the hot streamer; the arrow directed to the negative $x$ axis is $%
\mathbf{\nabla }n_{0}$ of the equilibrium density.

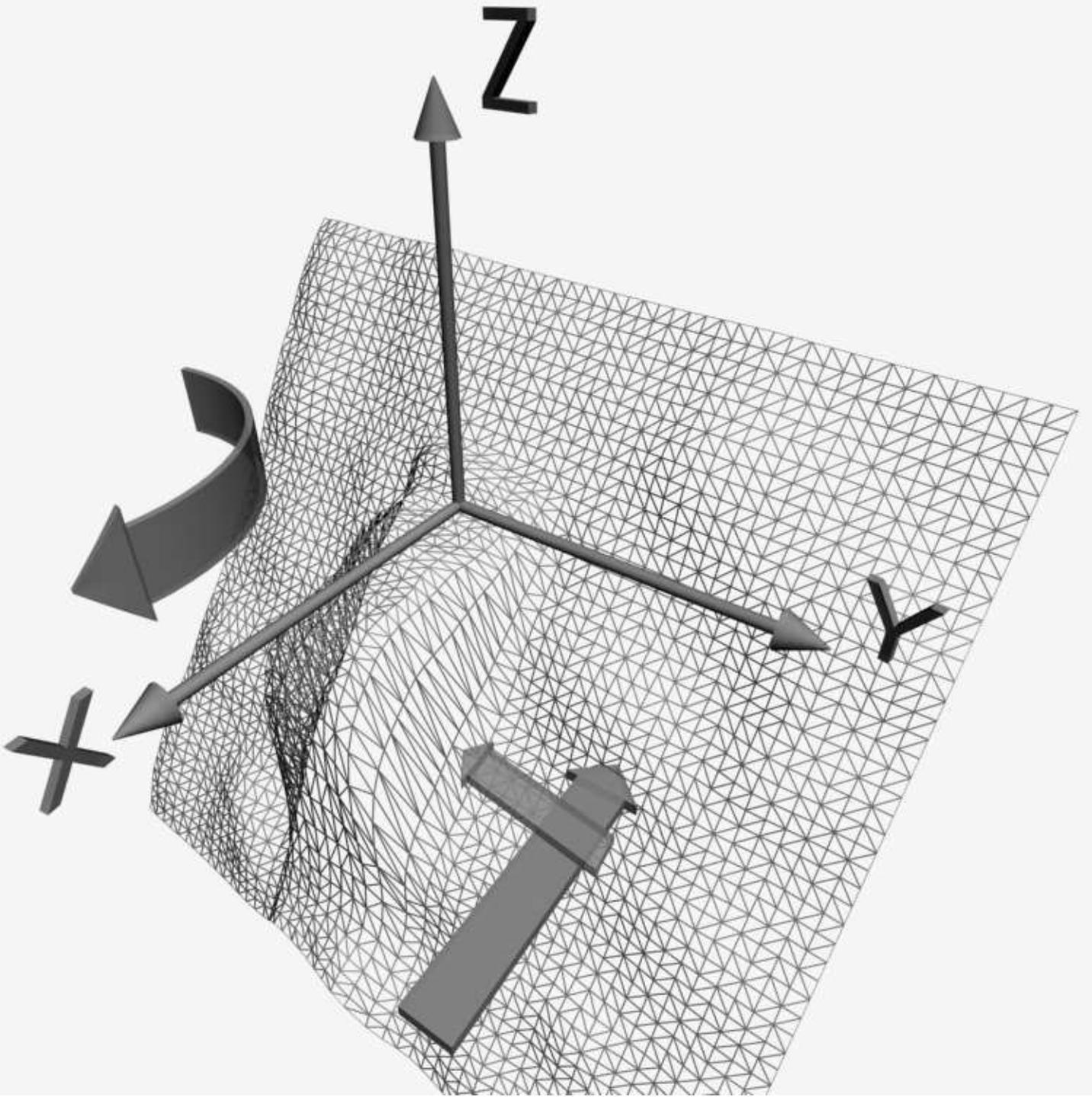
\end{document}